\documentclass[journal]{IEEEtran}
\usepackage{cite}
\usepackage{fleqn}
\usepackage{eufrak}
\usepackage{amsfonts}
\usepackage{amsmath,amssymb,amsfonts,amsmath,amsthm}
\usepackage{algorithm,algorithmic}
\allowdisplaybreaks
\usepackage{graphicx}
\usepackage{textcomp}
\usepackage{bm}
\usepackage{stfloats}
\usepackage{indentfirst} 
\usepackage{hyperref}
\usepackage{color}
\usepackage{url}
\usepackage{array}
\newcolumntype{I}{!{\vrule width 3pt}}
\newlength\savedwidth

\newlength\savewidth
\newcommand\shline{\noalign{\global\savewidth\arrayrulewidth
                           \global\arrayrulewidth 1.5pt}%
                  \hline
                  \noalign{\global\arrayrulewidth\savewidth}}

\newcommand{\RNum}[1]{\uppercase\expandafter{\romannumeral #1\relax}}
\hypersetup{colorlinks=true, linkcolor=black,filecolor=gray,urlcolor=black,citecolor=black,}

\ifCLASSINFOpdf
\else  
\fi

\begin{document}

\title{An Application-Driven Non-Orthogonal Multiple Access Enabled Computation Offloading Scheme}
%
%
% author names and IEEE memberships
% note positions of commas and nonbreaking spaces ( ~ ) LaTeX will not break
% a structure at a ~ so this keeps an author's name from being broken across
% two lines.
% use \thanks{} to gain access to the first footnote area
% a separate \thanks must be used for each paragraph as LaTeX2e's \thanks
% was not built to handle multiple paragraphs
%~\IEEEmembership{Student Member,~IEEE,}

\author{Qiqi~Ren,
        Jian~Chen,~\IEEEmembership{Member,~IEEE,}
        Omid~Abbasi,~\IEEEmembership{Graduate Student Member,~IEEE,}
        Gunes~Karabulut~Kurt,~\IEEEmembership{Senior Member,~IEEE,}
        Halim~Yanikomeroglu,~\IEEEmembership{Fellow,~IEEE,} 
        and F.~Richard~Yu,~\IEEEmembership{Fellow,~IEEE}% <-this % stops a space
%~\IEEEmembership{Student Member,~IEEE,}
\thanks{This work is supported by the National Natural Foundation of
China (Grant No. 61771366 and Grant No. 61901312) (Corresponding Author: Jian Chen).}% <-this % stops a space
\thanks{Q. Ren and J. Chen are with the State Key Laboratory of Integrated Service Networks, Xidian University, Xi' an 710071, Shaanxi, P. R. China, (email: renqiqi5277@gmail.com; jianchen@mail.xidian.edu.cn).}% <-this % stops a space
\thanks{O. Abbasi, H. Yanikomeroglu, and F. R. Yu are with the Department of Systems and Computer Engineering, Carleton University, Ottawa, ON, Canada (email: \{omidabbasi, halim\}@sce.carleton.ca; richard.yu@carleton.ca). }% <-this % stops a space
\thanks{G. K. Kurt is with the Department of Communications and Electronics Engineering, Istanbul Technical University, 34469, Istanbul, Turkey, e-mail: gkurt@itu.edu.tr.  }
\thanks{Copyright (c) 20xx IEEE. Personal use of this material is permitted. However, permission to use this material for any other purposes must be obtained from the IEEE by sending a request to pubs-permissions@ieee.org.}}

% The paper headers
\markboth{Journal of \LaTeX\ Class Files,~Vol.~14, No.~8, August~2015}%
{Shell \MakeLowercase{\textit{et al.}}: An Application-Driven Non-Orthogonal Multiple Access Enabled Computation Offloading Scheme}

% make the title area
\maketitle

% As a general rule, do not put math, special symbols or citations
% in the abstract or keywords.
\begin{abstract}
To cope with the unprecedented surge in demand for data computing for the applications, the promising concept of multi-access edge computing (MEC) has been proposed to enable the network edges to provide closer data processing for mobile devices (MDs). Since enormous workloads need to be migrated, and MDs always remain resource-constrained, data offloading from devices to the MEC server will inevitably require more efficient transmission designs. The integration of non-orthogonal multiple access (NOMA) technique with MEC has been shown to provide applications with lower latency and higher energy efficiency. However, existing designs of this type have mainly focused on the transmission technique, which is still insufficient. To further advance offloading performance, in this work, we propose an application-driven NOMA enabled computation offloading scheme by exploring the characteristics of applications, where the common data of the application is offloaded through multi-device cooperation. Under the premise of successfully offloading the common data, we formulate the problem as the maximization of individual offloading throughput, where the time allocation and power control are jointly optimized. By using the successive convex approximation (SCA) method, the formulated problem can be iteratively solved. Simulation results demonstrate the convergence of our method and the effectiveness of the proposed scheme.  
\end{abstract}

% Note that keywords are not normally used for peerreview papers.

% For peer review papers, you can put extra information on the cover
% page as needed:
% \ifCLASSOPTIONpeerreview
% \begin{center} \bfseries EDICS Category: 3-BBND \end{center}
% \fi
%
% For peerreview papers, this IEEEtran command inserts a page break and
% creates the second title. It will be ignored for other modes.
\IEEEpeerreviewmaketitle

\section{Introduction}
% The very first letter is a 2 line initial drop letter followed
% by the rest of the first word in caps.
% 
% form to use if the first word consists of a single letter:
% \IEEEPARstart{A}{demo} file is ....
% 
% form to use if you need the single drop letter followed by
% normal text (unknown if ever used by the IEEE):
% \IEEEPARstart{A}{}demo file is ....
% 
% Some journals put the first two words in caps:
% \IEEEPARstart{T}{his demo} file is ....
% 
% Here we have the typical use of a "T" for an initial drop letter
% and "HIS" in caps to complete the first word.
\subsection{Background}
 \IEEEPARstart{T}{he} booming development of the Internet of Things (IoT) drives the process of smart cities and Industry 4.0 \cite{b58}. It brings physical items into the digital field and enhances the acquisition of physical information, which makes it possible to understand or visualize the physical world in scalable horizontal \cite{b60,b63}. Applications represented by extended reality (XR, including virtual, augmented, and mixed reality), intelligent transportation, etc., which can convert data-rich complex scenarios into accessible applications, bridge the gap between the real world and digital world and bring fine-grained and even seamless experience (e.g., data analysis, and virtual visualization) to the industry at the required level. Taking medical XR as an example, during the COVID-19 pandemic, detailed patient cases can be visualized in multi-department doctors by specialized  XR application and various collected physical information (e.g., physical indicators, and scanned computed tomography (CT) data) without leaving their offices for holding seminars, demonstrations, and simulated surgeries \cite{b61,b62}. Such applications make the understanding and visualization easier, which can improve the communication and decision-making efficiency of participants. 
 
The applications with fine-grained seamless experience require a large amount of computation within the critical latency constraints, including the analysis and understanding of physical information, rendering virtual environment, step-by-step instruction superimposition, and overlaying design components to existing modules. Therefore, they place high demands on computation and battery capacity of mobile devices (MDs). However, MDs are always resource-constrained. For example, the standalone XR headset, as a direction the market is taking, is designed not to be tethered to supporting devices (e.g. computers), which constrains its resources \cite{b16,b14}. In order to better support these applications, the emerging paradigm of multi-access edge computing (MEC) extends computing and networking resources to the edge of network\cite{b13}. By enabling MDs to migrate their complicated computing loads to a proximate server, MEC offers MDs the advantages of proximity, powerful energy supplementation, real-time radio network information, and strong computing capability---in short, a superior service experience \cite{b52}. 

Despite its advantages, MEC requires data to be offloaded between the MD and the MEC server (MS) over wireless links. Generally, computation offloading means transmitting data to the server through the network's uplink, which plays a key role in improving the implementation of applications. In the initial 5G deployment, best-effort services focusing on downlink data transmission were mainly actualized \cite{b53,b74}. However, in the evolution of 5G, uplink enhancements and technology to improve the offloading performance are more important for the higher demands of applications, particularly cases where large amounts of data need to be offloaded \cite{b19,b73}.

Recently, non-orthogonal multiple access (NOMA) has shown great potential in further satisfying the ever-increasing communication demands of future cellular networks. Unlike conventional orthogonal multiple access (OMA) schemes, where orthogonal time/bandwidth resource blocks are allocated to MDs, NOMA enables multiple MDs to share the same time and radio resources by adopting superposition coding (SC) at the transmitter, which promises to fundamentally change the design of multiple access technologies \cite{b35}. NOMA splits users or devices in power domain by exploiting efficient multi-user detection (MUD) techniques, such as successive interference cancellation (SIC) at the receiver, which mitigate multiple access interference. Compared with OMA, NOMA provides the flexibility of multiple connections, superior spectral- and energy- efficiency, and greater network throughput by making better utilization of scarce radio resources \cite{b23}. 

Motivated by the benefits of NOMA, applying NOMA to MEC can improve the transmission throughput between MDs and MS, which not only reduces the overall latency, but also reduces the energy consumption in computation offloading, therefore helping systems cope with the high demands of the applications. 

\subsection{Related Work}
In this subsection, we begin with a review of related literature on computation offloading. This is followed by a review of related research on NOMA and NOMA-enabled computation offloading.

\textit{Research on computation offloading.} Generally, from the perspective of applications, studies on computation offloading can be classified into three categories, based on the partibility and the dependence of its computation tasks. The first category is binary offloading, where each MD either offloads the entirety of its computation task to the server or performs that locally \cite{b25,b76,b26}. For example, the works \cite{b25} and \cite{b76} studied binary task offloading in wireless virtualized network and wireless blockchain network, respectively. The authors in \cite{b26} proposed a framework of offloading tasks from a single MD to multiple servers, which can be regared as offloading diversity. The second category is partial offloading, where each MD can offload a part of its computation task to the server while the rest of the task is executed locally \cite{b28,b29}. Wang \textit{et al.} in \cite{b28} designed an energy- and latency-optimal partial offloading strategy for dynamic voltage scaling MDs. Kiani \textit{et al.} in \cite{b29} proposed a hierarchical MEC architecture where computation resources were offered in an auction-based profit maximization manner. The third category is dependency-based offloading, where the input of the current task usually consists of the computed results of the previous tasks \cite{b31,b12}. In \cite{b31}, a dependency-based offloading scheme was studied, where the duplicated tasks were offloaded by the specific mobile user to reduce the energy consumption. The authors in \cite{b12} proposed a caching-enhanced strategy considering the dependency among tasks, where the popular computation results could be cached at the servers to reduce the execution delay.

\textit{Research on NOMA.} Due to its superior advantages, NOMA has attracted more research interests in the works \cite{b33,b34,b35,b36,b49,b51,b37,b38,b39,b41}. In \cite{b34}, Zhu \textit{et al.} provided the optimal downlink NOMA power allocation in a closed or semi-closed form to improve fairness, sum-rate and energy efficiency. A comprehensive sub-channel assignment, power allocation, user scheduling scheme in a downlink NOMA network was proposed in \cite{b35} to achieve a balance between fairness and sum-rate. Except the resource management in downlink NOMA networks, several works also focused on uplink ones \cite{b36,b49,b51}. For instance, the authors in \cite{b36} explored joint base station association and power control to maximize system-wide utility and minimize the total power consumption for uplink NOMA. A multi-carrier uplink NOMA system was investigated in \cite{b51}, which requires to select the subcarriers for each user and distribute the transmission power to maximize the energy efficiency. Furthermore, there are some studies on the benefits of integrating NOMA with wireless powered transfer, multi-input-multi-output (MIMO), cognitive radio and cooperative relay to improve the performance of system\cite{b37,b38,b39,b41}.

\textit{Research on NOMA-enabled computation offloading.}
To reduce energy consumption, Kiani \textit{et al.} in \cite{b42} proposed a framework which exploits the benefits of integrating NOMA in MEC. Also focusing on the shortage of energy, the authors in \cite{b46} designed a multi-antenna NOMA computation offloading scheme, where both partial and binary offloading were considered. The resource management in work \cite{b57} was performed more comprehensively with the joint optimization of transmit power, time allocation for offloading and downloading, and task partitions. To reduce latency, Ding \textit{et al.} in \cite{b44} proposed to establish standards to select different offloading modes among OMA, pure NOMA, and hybrid NOMA for MEC. The authors in \cite{b45} studied the latency fairness among devices by jointly optimizing SIC ordering and computation allocation. In multi-carrier NOMA with the MEC-assisted wireless virtualized network, the authors provided a flexible video delivery scheme to enhance the total revenue of the network slices \cite{b43}. To improve the sum computation rate, Zeng \textit{et al.} in \cite{b56} studied computation mode selection  in a NOMA-enabled wireless-powered MEC network. The benefits of integrating NOMA in MEC networks have been established in terms of energy consumption, latency, and throughput, etc. However, in terms of the offloading scheme design, most of the existing works have focused more on the transmission technique itself, giving little consideration to the characteristics of the applications. In this work, we plan to address this gap in the literatures. 

\subsection{Motivations and Contributions}
As shown in \cite{b42,b44} and \cite{b56}, NOMA-enabled computation offloading achieves better performance in terms of latency, energy consumption, and offloading throughput. However, the specific characteristics of applications should also be taken into account, because it can also improve offloading performance. As mentioned in the related works on MEC, the dependence of tasks should be considered and the partibility of tasks should be exploited. Actually, the tasks in the computational intensive application represented by XR have unique properties. On one hand, when different devices (users) run the same application, partial data is common and the results of that can be shared among devices, while other data is individual. For example, when multi-department doctors use mobile devices (e.g., XR headsets) to visualize the detailed patient cases for holding seminars, the tasks such as rendering virtual environment according to physical details of the patient are common, including rendering organ VR model, and analyzing physical indicators, while other tasks such as the holders' preferred instructions and designs are individual, including targeted therapies and simulated surgery observations from specific angles, etc. On the other hand, the individual tasks are usually executed on the basis of the computed results of the previous common tasks. For instance, in the same XR application, different devices may take different actions on the basis of the generated virtual environment. 

However, most of the existing offloading schemes handle tasks independently without distinguishing the common data and individual data, which may result in repeated offloading to MDs \cite{b25,b76,b26,b28,b29,b42,b46,b57,b44,b45,b43,b56}. This will undoubtedly affect the offloading performance of the subsequent individual/personal data because most MDs remain energy-constrained and most applications are delay-critical. Note that in order to avoid redundant common data offloading, although the well-conditioned device can be selected to offload the entirety of the common data, the fairness among devices in the common data offloading stage will be poor, due to the fact that all devices share the computed results while only a single device offloads the entirety of the common data \cite{b31}. In addition, for cases with large common data volumes, the selected device which offloads the entirety of common data will consume too many resources (i.e.,  time and energy), which may lead to insufficient resources remaining to support the performance of individual data offloading.

To address this problem, we investigate an application-driven offloading scheme in a NOMA-enabled MEC network. In our scheme, considering the dependency between common data and individual data, the offloading is carried out in two stages, where the individual data offloading is performed after first guaranteeing the offloading of the common data. In addition, utilizing the partibility of the tasks, we consider exploring the multi-device collaboration, in which each MD offloads a part of the common data based on its condition (channel quality and available energy) to MS cooperatively. In our scheme, not only is offloading performance improved by avoiding redundant offloading and integrating resources rationally from all devices, but the fairness among the devices for offloading the common data is also improved. 

The scheme we present in this paper poses two challenges for resource management that we address at the outset. First, since both the offloading of common data and individual data are done in NOMA, optimally scheduling the time allocation for each stage within the permitted time duration must be properly considered. Second, the power control in a two-stage NOMA network should be appropriately determined, since it not only affects the multiple access interference in the same stage, thereby affecting the power control of other MDs, but there is also an influence between the two stages, as each MD is energy-constrained in the permitted application duration. It is worth noting that the time allocation and power control are interrelated, which makes it difficult to solve the joint optimization problem.

%: for example, when the allocated time of the first stage is reduced, the power control of all MDs needs to be adjusted accordingly in order to meet the requirement of common data volumes while mitigating multi-access interference, so the power control of the second stage will also be affected correspondingly, and vice versa. 

%\footnote{The available energy for each MD in the permitted duration may come from different resources (e.g., solar, wind, and radio), orr be accumulated by prior duration.}.

The main contributions of this paper are summarized as follows:
\begin{enumerate}
\item An application-driven two-stage offloading framework is proposed in a NOMA-enabled MEC network, where multi-device collaboration was utilized to offload the common data. By guaranteeing the common data offloading, we focus on the individual data offloading performance of each MD, which is more practical since each MD wants to offload data as much as possible.  
\item We formulate the optimization problem as the maximization of the minimal individual data offloading throughput. Subsequently, we propose a resource management scheme, where the time allocation for two NOMA stages and the power control for each MD in the two stages are jointly optimized.
\item According to the max-min structure of the individual throughput, the original non-convex optimization problem is converted into an equivalent optimization problem by introducing the energy variables and a series of slack variables. This can then be solved iteratively by using the successive convex approximation (SCA) method.
\end{enumerate}

The remainder of this paper is organized as follows. Section \RNum{2} describes the system model. In Section \RNum{3}, the problem formulation and solution are provided. Section \RNum{4} shows the simulation results of the system performance. Finally, Section \RNum{5} concludes this paper.

\section{System Model}
We consider a multi-device MEC-assisted computation offloading system, consisting of $N$ mobile devices (MDs), labeled as $\mathcal{N}=\{1,2,\dots,N\}$, and 
an MEC server (MS). Both MDs and MS are equipped with a single antenna, and the MDs are energy-constrained devices. The MS, as the controller of the system, is equipped with a communication module that periodically broadcasts probe requests to collect MDs' feedback, including knowledge of channel state information (CSI), available energy, and key parameters of the application to be executed, and broadcasts the offloading decision after calculating the optimal resource allocation based on the information collected\cite{b5}.  We assume that the MDs in this system execute an application that contains two dependent parts: a common data part and an individual data part, which may correspond to the VR model rendering for organs and the targeted therapies in practical medical situations, respectively. The system operates in two stages. At first, all MDs cooperatively offload common data to the MS, with each MD offloading a part of the common data and then the MS computing the common data. Next, after obtaining the computed results of the common data from the MS, each MD performs the individual data offloading process. In order to design a targeted offloading scheme, in this paper, we make the following assumption:

%\textbf{Assumption 1:} \textit{The consumption of energy and time at MS are omitted.} 
%
%This assumption has also been applied in many works due to the following  reasons \cite{b24,b54}: Compared with MDs, MS has more powerful computation ability and a more stable energy supplement, so the consumptions from computation and delivery of computed results at MS can be neglected. Besides, the data of computed results are always several bits, which is far less than the offloading data, and thus the downlink results delivery can be ignored. With this assumption, we can focus on the data offloading process and design an effective offloading scheme.
   %such as the objective recognition results, 

\textbf{Assumption 1:} \textit{Each MD always tends to offload individual data as much as possible on the  premise of successful common data offloading.} 

Due to their limited energy and computing capabilities, MDs cannot handle computationally intensive, latency-critical, and energy-consuming applications alone. Besides, these devices are in direct contact with the body, which need to keep cool, and running energy-consuming data will certainly lead to high temperatures. With the help of MEC, whose server is located close to MDs with superior computation capability and equipped with constant energy supply, MDs can obtain a more immediate computational response. Accordingly, each MD in this system tends to offload its workloads to the MS. Concretely, considering the dependency between common data and individual data, and to avoid redundant offloading, common data need to be processed first. Each MD then uses the remaining resources to offload individual data as much as possible. Note that offloading decisions for common data and individual data exploit task partibility.\footnote{Each MD is responsible for part of the common data and then offload individual data as much as possible. The offloaded data amount of each MD, also considered as task splitting results, can be obtained by calculating the data throughput that each MD can achieve at each stage based on the resource optimization results, as detailed later in this section.} Based on the above analysis, this paper focuses on how to improve the offloading performance for resource-constrained mobile devices so that the holders can experience better seamless performing and wearing/contacting comfort under the limited available resources of MDs.  \\
\begin{figure}[!t]
%\centerline{\includegraphics[width=\columnwidth]{s.pdf}}
\centerline{\includegraphics[width=9cm]{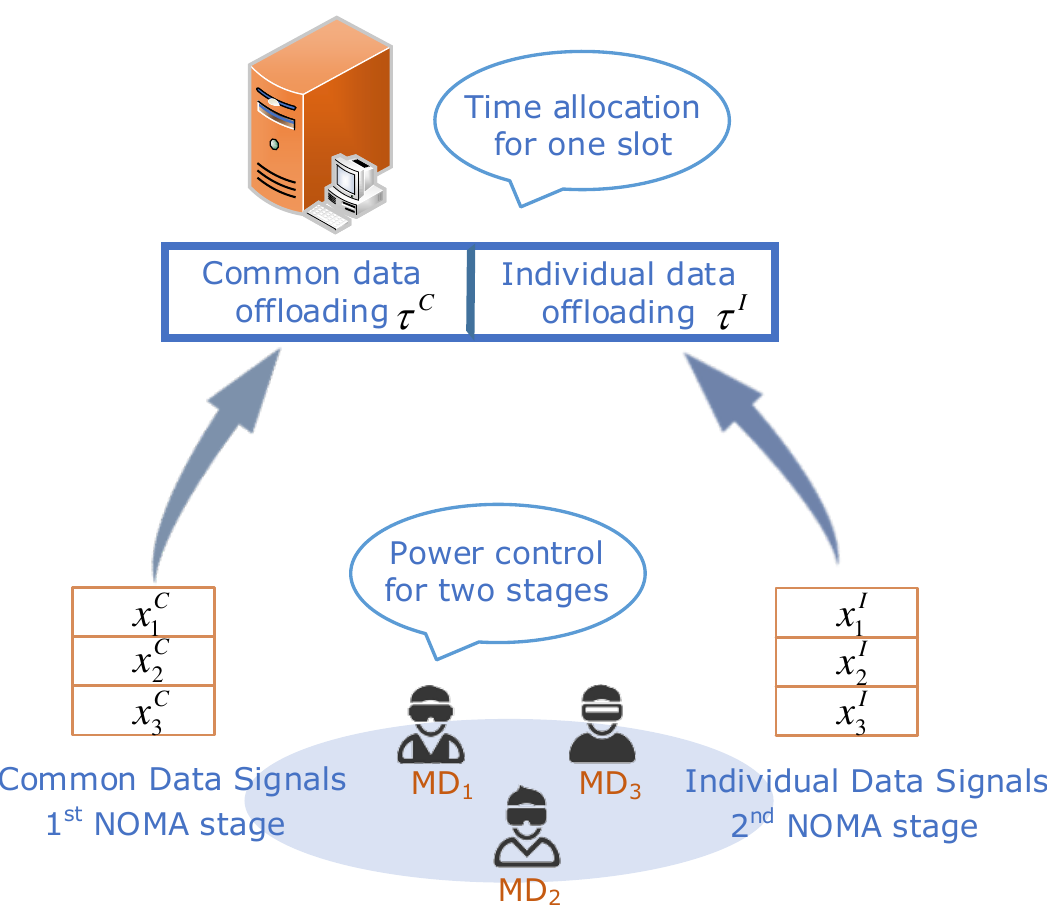}}
\caption{An example of a two-stage NOMA-enabled computation offloading system.}
\label{fig1}
\end{figure} 
\indent By employing NOMA, all MDs share the total frequency band $W$ for the computation offloading of one application, and the MS is equipped with a successive
interference cancellation (SIC) receiver to perform decoding. Consequently, the computation offloading of the application is processed in a two-stage NOMA uplink wireless transmission mode within a time slot. As the example in Fig. 1 shows, the MS
divides the time slot into two sub-slots, denoted by $\tau ^C$ and $\tau ^I$, corresponding to the common data offloading and individual data offloading, respectively. In the first sub-slot, all MDs contribute to offloading common data to the MS cooperatively, where each MD offloads a part of the common data. After that, each MD offloads its individual data in the second sub-slot. Both the common and individual data offloading are performed with NOMA.   

We consider a block fading channel model, which means the channel parameters remain unchanged in the period of one time slot, but vary independently from one slot to another. We use $|h_n|^2$ to denote the channel gain from the $n$-th MD to the MS, which is defined by $|h_n|^2=|g_n|^2L(d_n)$, where $g_n\sim\mathcal{CN}(0,1)$ is the Rayleigh fading channel coefficient from the $n$-th MD to the MS, $L(d_n)=d_n^{-\alpha}$ is the large-scale channel pathloss coefficient, $d_n$ is the distance between the $n$-th MD and the MS, and $\alpha$ is the pathloss exponent. Without loss of generality, we assume that all MDs are sorted in descending order and can be expressed as 
$|h_1|^2\ge |h_2|^2\ge\dots\ge|h_N|^2,\quad \forall n\in\mathcal{N}$. According to this order, the SIC receiver of the MS decodes the signal from the given MD by treating the signals from the subsequent MDs as interference in both of the two NOMA stages. 

In the first stage, all MDs offload common data to the MS in a collaborative manner in sub-slot $\tau^C$, where each MD contributes partially to the common data offloading and different power levels will be optimally determined. Let $\pmb{P^C}=\{P_n^C, n\in\mathcal{N}\}$, in which $P_n^C$ denotes the transmitted power for the link from the $n$-th MD to the MS within the allocated time $\tau^C$. Then, the received common data signal of the MS from the $n$-th MD in the first stage is defined by
\begin{equation}\small\label{signal1}
y_n^C = \sqrt {P_n^C} {h_n}x_n^C + \sum\limits_{k = n + 1}^N {\sqrt {P_k^C} {h_k}x_k^C}  + {w^C},\quad \forall n\in\{1,\ldots,N-1\},
\end{equation}
where the first term of the right hand is the desired signal from the $n$-th MD in the first stage, in which $x_n^C$ denotes the transmitted partial common data signal by the $n$-th MD; the second term is the interference caused by other partial common data signals offloaded by the other MDs subsequently in the same frequency band; the last term $w^C$ is the additive white Gaussian noise (AWGN) during the first stage with zero mean and variance $\sigma^2$. It is noted that the signal from the $N$-th MD does not suffer from the interference, i.e., $y_{_N}^C = \sqrt {P_{_N}^C} {h_{_N}}x_{_N}^C  + {w^C}$. Thus, the common data offloading throughput from the $n$-th MD to the MS is given by
\begin{small}
\begin{equation}\label{rates}
\begin{split}
 R_n^C(\tau^C,\pmb{P^C}) ={\tau ^C}W{\log _2}\left(1 + \frac{{P_n^C}{\gamma _n}}{{1 + {\sum\limits_{k = n + 1}^N {P_k^C}{\gamma _k} }}}\right),\\
\quad\quad\quad\quad\quad\quad\quad\quad \forall n\in\{1,\ldots,N-1\},
\end{split}
\end{equation}
\end{small}
and
\begin{equation}\small\label{ratess}
\begin{split}
 R _{_N}^C(\tau^C,P _{_N}^C) ={\tau ^C}W{\log _2}\left(1 + {P_{_N}^C}{\gamma _{_N}}\right),
\end{split}
\end{equation}
where ${\gamma _n} = \frac{{{{\left| {{h_n}} \right|}^2}}}{{{\sigma ^2}}}$ denotes the normalized channel gain for the link from the $n$-th MD to the MS. It is noted that since all MDs cooperatively contribute to offloading the common data, the sum throughput in this stage needs to satisfy the corresponding common data amount requirement, which will be given in the next section. The sum throughput of received common data from all MDs to the MS is calculated by 
\begin{small}
\begin{equation}\label{sumrates}
 {R^C}(\tau^C,\pmb{P^C}) = \sum\limits_{n = 1}^N {R_n^C}(\tau^C,\pmb{P^C})  = {\tau ^C}W{\log _2}\left(1 + \sum\limits_{n = 1}^N {P_n^C{\gamma _n}}\right).
\end{equation}
\end{small}

Similarly, in the second stage, each MD transmits its individual data to be computed within sub-slot $\tau^I$, and the power levels for different MDs will also be optimally determined. Let $\pmb{P^I}=\{P_n^I, n\in\mathcal{N}\}$, where $P_n^I$ denotes the transmitted power for the link from the $n$-th MD to the MS within the allocated time $\tau^I$. Therefore, the received individual data signal of the MS from the $n$-th MD is defined by
\begin{equation}\small\label{signal2}
y_n^I = \sqrt {P_n^I} {h_n}x_n^I + \sum\limits_{k = n + 1}^N {\sqrt {P_k^I} {h_k}x_k^I}  + {w^I},
\end{equation}
where the first term of the right hand is the desired signal from the $n$-th MD in the second stage, in which $x_n^I$ denotes the transmitted individual data signal of the $n$-th MD; the second term  is the interference caused by other individual data signals transmitted by the other MDs subsequently in the same frequency band; the last term $w^I$ is the AWGN during the second stage with zero mean and variance $\sigma^2$. Similarly, the signal from the $N$-th MD does not suffer from the interference in this stage, i.e., $y_{_N}^I = \sqrt {P_{_N}^I} {h_{_N}}x_{_N}^I + {w^I}$. Afterwards, the individual data offloading throughput from the $n$-th MD to the MS is given by
\begin{small}
\begin{equation}\label{ratei}
\begin{split}
 R_n^I(\tau^I,\pmb{P^I}) ={\tau ^I}W{\log _2}\left(1 + \frac{{P_n^I}{\gamma _n}}{{1 + {\sum\limits_{k = n + 1}^N {P_k^I}{\gamma _k} }}}\right),\\
\quad \quad \quad \quad \quad \quad \quad \forall n\in\{1,\ldots,N-1\},
\end{split}
\end{equation}
\end{small}
%&= {\tau ^I}W{\rm{lo}}{{\rm{g}}_2}\left(1 + \frac{{P_n^I{{\left| {{h_n}} \right|}^2}}}{{{\sigma ^{\rm{2}}} + \sum\limits_{k = n + {\rm{1}}}^N {P_k^I{{\left| {{h_k}} \right|}^2}} }}\right)\\
and 
\begin{small}
\begin{equation}\label{rateii}
\begin{split}
 R _{_N}^I(\tau^I,P _{_N}^I) ={\tau ^I}W{\log _2}\left(1 + {P _{_N}^I}{\gamma _{_N}}\right).
\end{split}
\end{equation}
\end{small}
\section{Problem Formulation and Solution}

In this section, the joint time allocation and power control in the two-stage NOMA offloading network are investigated. On the premise of ensuring common data offloading, we describe the optimization problem as an individual data offloading throughput max-min problem to improve the worst individual offloading performance of all MDs\footnote{ Optimizing the worst offloading performance of all MDs can provide a more similar experience for the MDs, which means the experience of all MDs in the same application will not be affected by the worst MD. When the objective performance is improved, the amount of data that can be processed will also be increased and the workloads that need to be computed at MDs will be alleviated, so that the contacting/wearing comfort will not be affected and the perceived experience can be enhanced due to the less occurrence of task timeout. }. After mathematical transformations, we show that the proposed problem can be solved by exploiting the successive convex approximation (SCA) method. 
\subsection{Problem Formulation}

Following Assumption 1, all MDs tend to maximize their individual offloading performance as much as possible. Therefore, our objective is to maximize the minimal throughput among all MDs in the second stage by setting the variables $\tau^C$, $\tau^I$, $\pmb{P^C}$, and $\pmb{P^I}$ under some constraints. To this end, the optimization problem can be formulated as
%\begin{small} 
\begin{subequations}\small\label{op1}
\begin{align}
%\begin{equation}
&\max_ {\substack{{\tau}^{ C},{\tau}^{ I},\\ \pmb{P^{ C}},\pmb{P^{ I}}}}\ \min_{n\in{\mathcal N}} \quad R_n^I \vspace{-10pt} \\
%\end{equation}
%
\rm{s.t.}\quad &{\tau}^{ C}{P^{ C}_n}+{\tau}^{ I}{P^{ I}_n}\leq E_{n,\max},\quad\forall n,\label{op1b}\\
&{\tau ^{ C}} + {\tau ^{ I}} \leq {T_{\max }},\label{op1c}\\
&{\tau ^{ C}}W{\log _2}\left(1 + {{\sum\limits_{n = 1}^N }{P_n^{ C}} {\gamma _n}}\right) \geq K,\label{op1d}\\
&{\tau ^{ C}} > 0,\quad {\tau ^{ I}} > 0,\label{op1e}\\
&P_n^{ C} \ge 0,\quad P_n^{ I} \ge 0,\quad\forall n.\label{op1f}
\end{align}
\end{subequations}
%\end{small}
In problem \eqref{op1}, \eqref{op1b} constrains the energy consumption for the $n$-th MD to $E_{n,\max}$, where $n \in{\mathcal N}$. \eqref{op1c} denotes the time constraint of the application, where $T_{\max}$ is the tolerable offloading latency and can be regarded as a time slot. \eqref{op1d} guarantees the total throughput requirement during the common data offloading stage, where $K$ represents the required bits of the common data to be offloaded, which is determined by the specialized module of the application. \eqref{op1e} and \eqref{op1f} indicate the basic constraints that the variables should satisfy.

Problem \eqref{op1} is a non-convex optimization problem due to its non-smooth objective function, the existence of interference in the objective function, and the product relations between the variables in the objective function and constraints \eqref{op1b} and \eqref{op1d}. Generally, there is no standard method to solve problem \eqref{op1} efficiently due to its non-convexity. Therefore, in the following subsection, we transform the original problem into a concave problem step by step, so that the resource allocation can be developed more efficiently. 

\subsection{Problem Transformation}
In this subsection, we show how to transform the original optimization problem. First, we introduce an auxiliary variable $\phi$ to transform the original non-smooth problem into a smooth one. Then, Problem \eqref{op1} can be equivalently rewritten as
%\setlength{\mathindent}{0cm}
%\begin{small}
\begin{subequations}\small\label{op2}
\begin{align}
&\max_ {\substack{\phi,\;{\tau}^{ C},{\tau}^{ I},\\ \pmb{P^{ C}},\pmb{P^{ I}}}}\,\phi \vspace{-10pt} \\
\rm{s.t.}&{\tau ^{ I}}W{\log _2}\left(1 + \frac{{P_n^{ I}}{\gamma _n}}{{1 + {\sum\limits_{k = n + 1}^N {P_k^{ I}}{\gamma _k} }}}\right) \ge \phi ,\forall n\in\{1,\ldots,N-1\},\label{op2b}\\ \vspace{4ex}
&{\tau ^{ I}}W{\log _2}\left(1 + {P _{_N}^{ I}}{\gamma  _{_N}}\right) \ge \phi , \label{op2c}\\
&(\ref{op1b}),(\ref{op1c}),(\ref{op1d}),(\ref{op1e}),(\ref{op1f}).
\end{align}
\end{subequations}
%\end{small}
Second, to tackle the issue caused by the products of optimization variables in constraints \eqref{op1b}, \eqref{op1d}, \eqref{op2b} and \eqref{op2c}, we introduce an additional series of variables, i.e., $E_n^C=P_n^C\tau^C$ and $E_n^I=P_n^I\tau^I$, $\forall n$, which can be thought of as the real energy consumption of the $n$-th MD in the two stages. By substituting $\small{P_n^C = \frac{{E_n^C}}{{{\tau ^C}}}}$ and $\small{P_n^I = \frac{{E_n^I}}{{{\tau ^I}}}}$ in problem \eqref{op2}, we can obtain the following optimization problem: 
\begin{subequations}\small\label{op3}
\begin{align}
%\begin{equation}
&\max_ {\substack{\phi,\;{\tau}^{ C},{\tau}^{ I},\\ \pmb{E^{ C}},\pmb{E^{ I}}}}\,\phi \vspace{-10pt} \\
%\end{equation}
\rm{s.t.}\quad&E_n^{ C} + E_n^{ I} \le {E_{n,\max }},\quad\forall n\in\mathcal{N}, \label{op3b}\\
&{\tau ^{ C}}W\log_2 \left(1 + {\frac{\sum\limits_{n = 1}^N {E_n^{ C}{\gamma _n}}}{{{\tau ^{ C}}}}}  \right) \ge K, \label{op3c}\\
&{\tau ^{ I}}W{\log _2}  \left(1 + \frac{{\frac{{E_n^{ I}}{\gamma _n}}{{{\tau ^{ I}}}}}}{{1 +  {\frac{\sum\limits_{k = n + 1}^N{E_k^{ I}}{\gamma _k}}{{{\tau ^{ I}}}}} }} \right) \ge \phi ,\nonumber\\
&\quad\quad\quad\quad\quad\quad\quad\quad\quad\quad\forall n\in\{1,\ldots,N-1\},\label{op3d}\\ 
&{\tau ^{ I}}W{\log _2}  \left(1 + {\frac{{E _{_N}^{ I}}{\gamma  _{_N}}}{{{\tau ^{ I}}}}} \right) \ge \phi ,\label{op3f}\\
&E_n^{ C} \ge 0,\quad E_n^{ I} \ge 0,\quad\forall n\in\mathcal{N}, \label{op3e}\\
&(\ref{op1c}),(\ref{op1e}).
\end{align}
\end{subequations}
However, solving problem \eqref{op3} is still difficult due to the complicated expression in constraint \eqref{op3d}. Using the property of the logarithmic function, we therefore expand the throughput in the second stage for any $n\in\{1,\ldots,N-1\}$ as follows:
\begin{equation}\small\label{R4}
\begin{split}
&\mathord{\buildrel{\lower3pt\hbox{$\scriptscriptstyle\frown$}} 
\over R} _{\rm{n}}^I(\tau^I,\pmb{E^I})\\
 =&{\tau ^{ I}}W{\log _2}\left(1 + \frac{{ \frac{{E_n^{ I}{\gamma _n}}}{{{\tau ^{ I}}}}}}{{1 + {\frac{\sum\limits_{k = n + 1}^N {E_k^{ I}{\gamma _k}}}{{{\tau ^{ I}}}}} }}\right)\\
=&{\tau ^{ I}}W{\log _2}\left(1 + { \frac{\sum\limits_{j = n}^N {E_j^{ I}}{\gamma _j}}{{{\tau ^{ I}}}}}\right ) -{\tau ^{ I}}W{\log _2}\left(1 + { \frac{\sum\limits_{k = n+1}^N {E_k^{ I}}{\gamma _k}}{{{\tau ^{ I}}}}}\right ).\\
\end{split}
\end{equation}
In equation (11), $\mathord{\buildrel{\lower3pt\hbox{$\scriptscriptstyle\frown$}}
\over R}^I_n$ represents the left hand side of constraint (10d). After this conversion, problem \eqref{op3} can be given by
\begin{small}
\begin{subequations}\label{op4}
\begin{align}
%\begin{equation}
&\max_ {\substack{\phi,\;{\tau}^{ C},{\tau}^{ I},\\ \pmb{E^{ C}},\pmb{E^{ I}}}}\,\phi \vspace{-10pt} \\
%\end{equation}
\rm{s.t.}\ &{\tau ^{ I}}W{\log _2}\Bigg(1 + { \frac{\sum\limits_{j = n}^N {E_j^{ I}}{\gamma _j}}{{{\tau ^{ I}}}}}\Bigg ) -{\tau ^{ I}}W{\log _2}\Bigg(1 + { \frac{\sum\limits_{k = n+1}^N {E_k^{ I}}{\gamma _k}}{{{\tau ^{ I}}}}}\Bigg )\nonumber\\ 
&\quad\quad\quad\quad\quad\quad\quad\quad\ge \phi,\forall n\in\{1,\ldots,N-1\}, \label{op4b}\\
&(\ref{op1c}),(\ref{op1e}),(\ref{op3b}),(\ref{op3c}),(\ref{op3f}),(\ref{op3e}).
\end{align}
\end{subequations}
\end{small}

\textbf{Remark 1:} It is worth noting that both of the two terms in the left hand of constraint \eqref{op4b} as well as the term in the left hand of constraint \eqref{op3c} have a similar form, which can be expressed by $z(\alpha,\pmb{\beta})=\alpha\log_2(1+{ \frac{\sum\limits_j {\beta_j}{\gamma _j}}{{{\alpha }}}})$. It is difficult to analyze the concavity of this form based on its Hessian Matrix which is in terms of the variable $\alpha$ and a set of variables $\pmb{\beta}=\{\beta_j,\forall j \in \mathcal{N}\}$. 
%In general, if the concavity of the partial terms cannot be guaranteed, let alone the whole constraint. 

Hence, similar to \cite{b50}, we also introduce the slack variables $S_1=\sum\limits_{n = 1}^N {E_n^C} {\gamma _n}$, 
${S_{2,n}} = \sum\limits_{j = n}^N {E_j^I} {\gamma _j}$, $\forall n\in\{1,\ldots,N\}$, and ${S_{3,n}} = \sum\limits_{k = n + 1}^N {E_k^I} {\gamma _k}$, $\forall n\in\{1,\ldots,N-1\}$, respectively, to tackle the concerns mentioned above. Following this, problem $\eqref{op3}$ can be reformulated as
\begin{subequations}\small\label{op5}
\begin{align}
%\begin{equation}
&\max_ {\substack{\phi,\;{\tau}^{ C},{\tau}^{ I},\\ \pmb{E^{ C}},\pmb{E^{ I}},\\S_1,\pmb{S_2},\pmb{S_3}}}\,\phi \vspace{-10pt} \\
%\end{equation}
\rm{s.t.}\quad&{\tau ^{ C}}W\log_2 \left(1 + {\frac{S_1}{{{\tau ^{ C}}}}}  \right) \ge K ,\label{op5b}\\
&{\tau ^{ I}}W{\log _2}\left(1 + { \frac{S_{2,n}}{{{\tau ^{ I}}}}}\right ) -{\tau ^{ I}}W{\log _2}\left(1 + { \frac{S_{3,n}}{{{\tau ^{ I}}}}}\right )\ge \phi ,\ \nonumber\\
&\quad\quad\quad\quad\quad\quad\quad\quad\quad\quad\quad\forall n\in\{1,\ldots,N-1\}, \label{op5c}\\
&{\tau ^{ I}}W{\log _2}\left(1 + { \frac{S_{2,N}}{{{\tau ^{ I}}}}}\right ) \ge \phi ,\label{op5g}\\
&{S_1} \le \sum\limits_{n = 1}^N {E_n^C{\gamma _n}},\label{op5d}\\
&{S_{2,n}} \le \sum\limits_{j = n}^N {E_j^I{\gamma _j}}, \quad \forall n\in\mathcal{N},\label{op5e}\\
&{S_{3,n}} \ge \sum\limits_{k = n + 1}^N {E_k^I{\gamma _k}} , \quad \forall n\in\{1,\ldots,N-1\},\label{op5f}\\
&(\ref{op1c}),(\ref{op1e}),(\ref{op3b}),(\ref{op3e}).
\end{align}
\end{subequations}

\textbf{Proposition 1:} Without loss of optimality to problem \eqref{op5}, the constraints of \eqref{op5d}, \eqref{op5e} and \eqref{op5f} will be met with equality. 
%which means that problem \eqref{op5} is equivalent to problem \eqref{op4}. 

\begin{proof}
Please see Appendix A.
\end{proof}

With the proof of Proposition 1, the relaxations introduced by these slack variables are tight.

\textbf{Proposition 2:} The constraints \eqref{op5b} and \eqref{op5c} are jointly concave in variables $\tau^C$ and $S_1$, as well as $\tau^I$ and $S_{2,n}$, respectively, and the constraint \eqref{op5c} is in the form of difference of two concave functions (D.C.) in variables $\tau^I$, $\pmb{S_{2}}$ and $\pmb{S_{3}}$.

\begin{proof}
Please see Appendix B.
\end{proof}

With the proof of Proposition 2, we know that all independent terms in the reformulated problem \eqref{op5} are either concave or linear.   

However, we should mention that although the introduction of the slack variables above make the independent terms of the constraint function concave, the resulting set of constraints \eqref{op5c} is still non-convex owing to the D.C. form. A widely used method to handle this kind of problem is the SCA method \cite{b4,b5}. The key idea of the SCA method is to iteratively approximate the original non-convex/non-concave problem as multiple subproblems of locally optimal, which are a series of convex/concave versions of the original problem \cite{b5}. In this case, we first express the individual data offloading throughput function in constraint \eqref{op5c} in the D.C. form as
\begin{equation}\small\label{R5}
\begin{split}
g_n(\tau^I, S_{2,n}, S_{3,n})=g_{1,n}(\tau^I, S_{2,n})-g_{2,n}(\tau^I, S_{3,n}), 
\end{split}
\end{equation}
where $g_{1,n}(\tau^I, S_{2,n})$ and $g_{2,n}(\tau^I, S_{3,n})$ are two concave functions defined as follows:
\begin{equation}\small\label{g1}
\begin{split}
g_{1,n}(\tau^I, S_{2,n})={\tau ^{ I}}W{\log _2}\left(1 + { \dfrac{S_{2,n}}{{{\tau ^{ I}}}}}\right ),
\end{split}
\end{equation}
\begin{equation}\small\label{g2}
\begin{split}
g_{2,n}(\tau^I, S_{3,n})={\tau ^{ I}}W{\log _2}\left(1 + { \dfrac{S_{3,n}}{{{\tau ^{ I}}}}}\right ),
\end{split}
\end{equation}
for any MD $n\in\{1,\ldots,N-1\}$.

Then, define $\pmb{\mathcal{A}^m}=\{\mathcal{A}_n^m,\forall n\in\{1,\ldots,N-1\}\}$, wherein $\mathcal{A}_n^m=(\tau^I[m], S_{3,n}[m])$, as the given local point set in the $m$-th iteration. It will be recalled that any concave function is globally upper-bounded by its first-order Taylor expansion at any given point. For a function $q(a,b)$ which is jointly concave with respect to variables $a$ and $b$, by applying first-order Taylor expansion at the given local point $(a_0,b_0)$, we get $q(a,b)\le q(a_0,b_0)+\nabla q_a(a,b)(a-a_0)+\nabla q_b(a,b)(b-b_0)$, where $\nabla q_a(a,b)$ and $\nabla q_b(a,b)$ indicate the gradient of function $q(a,b)$ about variables $a$ and $b$, respectively. With this equation, we obtain the inequality in terms of the concave function $g_{2,n}(\tau^I, S_{3,n})$ at the given local point $\mathcal{A}_n^m$ as follows:
\begin{equation}\small\label{R7}
\begin{split}
&g_{2,n}(\tau^I, S_{3,n})\\
&\le B_n(\tau^I[m], S_{3,n}[m])+D_n\Big(\tau^I[m], S_{3,n}[m]\Big)\Big(\tau ^I-\tau ^I[m]\Big)\\
&\quad+Q_n\Big(\tau^I[m], S_{3,n}[m]\Big)\Big(S_{3,n}-S_{3,n}[m]\Big)\\
&\quad={\hat g_{2,n}}\Big(\tau^I,S_{3,n}\Big).
\end{split}
\end{equation}
In \eqref{R7}, $B_n\Big(\tau^I[m], S_{3,n}[m]\Big)$ represents the value of function $g_{2,n}$ at the given point $\mathcal{A}_n^m$, which is defined by  
\begin{equation}\small\label{B}
\begin{split}
B_n\Big(\tau^I[m], S_{3,n}[m]\Big)={\tau ^I}\left[m\right]W{\log _2}\left (1 + \frac{{{S_{3,n}}\left[m\right]}}{{{\tau ^I}\left[m\right]}}\right ),\\
\end{split}
\end{equation}
$D_n\Big(\tau^I[m], S_{3,n}[m]\Big)$ represents the gradient of $g_{2,n}$ related to variable $\tau^I$ at the given point $\mathcal{A}_n^m$, which is denoted by
\begin{equation}\small\label{D}
\begin{split}
&D_n\Big(\tau^I[m], S_{3,n}[m]\Big)\\
&=\frac{W}{{\ln 2}}\left(\ln \left(1 + \frac{{{S_{3,n}}\left[m\right]}}{{{\tau ^I}\left[m\right]}}\right) - \frac{{\frac{{{S_{3,n}}\left[m\right]}}{{{\tau ^I}\left[m\right]}}}}{{1 + \frac{{{S_{3,n}}\left[m\right]}}{{{\tau ^I}\left[m\right]}}}}\right),
\end{split}
\end{equation}
and $Q_n\Big(\tau^I[m], S_{3,n}[m]\Big)$ represents the gradient of $g_{2,n}$ related to variable $S_{3,n}$ at the given point $\mathcal{A}_n^m$, which is expressed by
\begin{equation}\small\label{Q}
\begin{split}
&Q_n\Big(\tau^I[m], S_{3,n}[m]\Big)=\frac{W}{{\ln 2 \cdot \left(1 + \frac{{{S_{3,n}}\left[m\right]}}{{{\tau ^I}\left[m\right]}}\right)}}\, .
\end{split}
\end{equation}
Accordingly, we can obtain the upper bound of $g_{2,n}(\tau^I, S_{3,n})$, which is denoted as ${\hat g_{2,n}} \Big(\tau^I,S_{3,n}\Big)$. From \eqref{R5} to \eqref{Q}, we know that
\begin{equation}\small\label{R8}
\begin{split}
&g_n(\tau^I, S_{2,n}, S_{3,n})\ge g_{1,n}(\tau^I, S_{2,n})-\hat g_{2,n}(\tau^I, S_{3,n}), \\
&\quad\quad\quad\quad\quad\quad\quad \quad\forall n\in\{1,\ldots,N-1\},
\end{split}
\end{equation}
the right hand of which is a concave function in terms of variables $\tau^I$ and $S_{3,n}$, owing to the concave property of $g_{1,n}$ and the linear property of the remaining terms.  
Note that with \eqref{R8}, the individual offloading data throughput function in constraint \eqref{op5c} can be lower bounded by $\small{g_{1,n}(\tau^I, S_{2,n})-\hat g_{2,n}(\tau^I, S_{3,n})}$. Therefore, with any given local point set $\pmb{\mathcal{A}^m}$ and the approximation by the lower bounds in \eqref{R8}, problem \eqref{op5} can be recast as follows:
\begin{subequations}\small\label{op6}
\begin{align}
&\max_ {\substack{\phi,\;{\tau}^{ C},{\tau}^{ I},\\ \pmb{E^{ C}},\pmb{E^{ I}},\\S_1,\pmb{S_2},\pmb{S_3}}}\,\phi \vspace{-10pt} \\
\rm{s.t.}\quad&g_{1,n}(\tau^I, S_{2,n})-\hat g_{2,n}(\tau^I, S_{3,n})\ge \phi ,\ \forall n\in\{1,\ldots,N-1\},\label{op6b}\\
&(\ref{op1c}),(\ref{op1e}),(\ref{op3b}),(\ref{op3e}),(\ref{op5b}),(\ref{op5g}),(\ref{op5d}),(\ref{op5e}),(\ref{op5f}).
\end{align}
\end{subequations}

Owing to the concave property of constraints \eqref{op5b}, \eqref{op5g}, and \eqref{op6b} and the linear property of the objective function and constraints \eqref{op1c}, \eqref{op1e}, \eqref{op3b}, \eqref{op3e}, \eqref{op5d}, \eqref{op5e} and \eqref{op5f}, problem \eqref{op6} is verified to be a standard concave optimization problem. By iteratively using linear functions to approximate $g_{2,n}$ around a series of given points, the resulting sets that satisfy the constraint \eqref{op6b} in a series of approximated problems are strict subsets of the set that satisfies non-concave constraint \eqref{op5c} in problem \eqref{op5}. This fact implies that the solution of each approximated problem \eqref{op6} is locally optimal for the problem \eqref{op5}. Furthermore, we will prove the convergence of the proposed approximation by the SCA method, which is provided in the following proposition. 

\textbf{Proposition 3:} By applying the SCA method to solve \eqref{op5}, the objective function value of the original problem is non-decreasing with iteration number $m$, i.e., $\phi[m]\ge\phi[m-1]$, and the resulting approximation is guaranteed to converge. 
\begin{proof}
Please see Appendix C.
\end{proof}
 
Note that although in Proposition 3 we prove the local optimality of the objective function with the help of the SCA method, the global optimality cannot be guaranteed due to the non-concavity of the original problem in \eqref{op4}.\\

%\subsection{Joint Time Allocation and Energy Partition}

Since problem \eqref{op6} is a standard convex optimization problem, we can find out its optimal solution by using CVX solver. Note that in the left hands of constraints \eqref{op5b}, \eqref{op6b}, and \eqref{op5g}, there exsist concave perspective function forms, i.e., $\small{f={\tau ^{ C}}W\log_2 \left(1 + {\frac{S_1}{{{\tau ^{ C}}}}} \right)}$, $\small{g_{1,n}={\tau ^{ I}}W{\log _2}\left(1 + { \frac{S_{2,n}}{{{\tau ^{ I}}}}}\right )}$ for $n\in\{1,\ldots,N-1\}$, and $\small{g_N=\tau ^{ I}W{\log _2}\left(1 + { \frac{S_{2,N}}{{{\tau ^{ I}}}}}\right )}$. However, there is no direct function can be used for perspective function in CVX tool box. By using convex function tool $rel\_entr(x,y)=x.*\ln(x/y)$, we can respectively construct them as follows:
\begin{equation}\small
{f=-\frac{W}{\ln2}*rel\_entr\left(\tau^C,\tau^C+S_1\right)},
\end{equation}
\begin{equation}\small
\begin{split}
&g_{1,n}=-\frac{W}{\ln2}*rel\_entr\left(\tau^I,\tau^I+S_{2,n}\right),\\
&\quad\quad\quad\quad\quad\quad\quad\quad\quad\forall n\in\{1,\ldots,N-1\},
\end{split}
\end{equation}
 and 
\begin{equation}\small
{g_N=-\frac{W}{\ln2}*rel\_entr\left(\tau^I,\tau^I+S_{2,N}\right)}.
\end{equation}
 Therefore, with these expressions, the optimal solution of problem \eqref{op6} can be directly solved by CVX.

The SCA iteration process is presented in Algorithm 1, where the non-concave function of constraint \eqref{op5c} is approximated by a series of concave functions around the given local point set, which is the partial optimal values obtained by solving problem \eqref{op6}. The total complexity of Algorithm 1 is at most $O({N_m}{N^3})$, where $N_m$ is the maximal number of iterations for the SCA method, and $O({N^3})$ is the complexity for solving problem \eqref{op6} \cite{b55}. Therefore, Algorithm 1 is of polynomial complexity.

\begin{algorithm}
\caption{The SCA method for solving problem \eqref{op5}.}
\label{alg:A}
%=\{\tau^I[0], \pmb{S_3}[0]\}
\begin{algorithmic}[1]
\STATE Initialize ${\varepsilon }$, $\pmb{\mathcal{A}^0}$, $N_m$ and set $m=0$.
\REPEAT 
\STATE Calculate $\pmb{B}$, $\pmb D$ and $\pmb Q$ around the given local point set $\pmb{\mathcal{A}^m}$, according to \eqref{B}--\eqref{Q}.
\STATE Obtain $\Lambda=\{\phi^*,{\tau}^{{ C}{*}},{\tau}^{ I*},$$\pmb{E^{ C}}^*,$$\pmb{E^{ I}}^*,{S_1}^*,\pmb{S_2}^*,\pmb{S_3}^*\}$ by solving convex problem \eqref{op6}.  
\STATE Record current $\Lambda$ as $\Lambda^\dag =\{\phi^\dag,{\tau}^{{ C}{\dag}},{\tau}^{ I\dag},\pmb{E^{ C}}^\dag,\pmb{E^{ I}}^\dag,{S_1}^\dag,\pmb{S_2}^\dag,\pmb{S_3}^\dag\}$. 
\STATE Save $\phi [m]=\phi^\dag$, update $\pmb{\mathcal{A}^{m+1}}$ by ${\tau}^{ I}[m+1]={\tau}^{ I\dag}$ and $\pmb{S_3}[m+1]=\pmb{S_3}^\dag$, and $m=m+1$. 
\UNTIL{$\phi[m]-\phi[m-1]\le \varepsilon$, or $m>N_m$}.
\end{algorithmic}
\end{algorithm} 
% with the objective of maximizing the minimal individual throughput
\section{Simulation Results and Discussions}

 In this section,we present simulation results to evaluate the performances of the following schemes: 
\begin{enumerate}
\item S-NOMA: This scheme select a single MD to offload the entirety of the common data \cite{b31}, and all MDs offload their individual data using NOMA.
\item S-OMA: This scheme select a single MD to offload the entirety of the common data \cite{b31}, and all MDs offload their individual data using OMA.
\item Benchmark: In this scheme, using NOMA, all MDs offload the common data altogether, and then they offload their individual data.
\item Proposed: In this scheme, using NOMA, all MDs cooperatively offload the common data, and then they offload their individual data.
\end{enumerate}

\begin{table}\nonumber
\caption{Simulation Parameters}
\begin{center}
\begin{tabular}{c|c|c}
\shline
Parameter & Definition & Value  \\ 
\hline
$W$ & Bandwidth & 1 MHz  \\ 
\hline
$N_0$ & Noise power spectral density & -174 dBm/Hz  \\
\hline
$\alpha$ & Pathloss factor & 3 \cite{b64} \\
\hline
$T_{\max}$ & Tolerable offloading latency & 1 s \cite{b54}  \\
\hline 
$K$ & Common data amount requirement & 1 Mbits to 12 Mbits  \\
\hline
$d$ & The distribution range of MDs & 200 m  \\
\hline
$E_{n,\max}$ & Available energy of MDs & 0.1 J to 0.3 J \\
\hline
$N_m$ & Maximal outer iteration indicator   & 50 \\
\hline
$\varepsilon$ & Convergence accuracy indicator   & $10^{-4}$ \\
\shline
\end{tabular}
\end{center}
\end{table}

In the simulations, we assume that in the MEC-assisted offloading system, there is an MS and  multiple MDs. The MDs are randomly distributed around the MS within certain range. The wireless channel gains is set as $|h|^2=|g|^2L(d)$, where $L(d)=d^{-\alpha}$ is large-scale channel pathloss coefficient, $d$ is the geographical distance between the MD and MS,  $\alpha$ is the pathloss exponent, and $g$ is small-scale fading coefficient following Rayleigh distribution with mean 0 and variance 1. The main simulation parameters are summarized in Table I. The Monte Carlo simulation is used for evaluating the performance of the proposed scheme. The simulation results are averaged over a large number of independent runs.
\begin{figure}[!t]
%\centerline{\includegraphics[width=\columnwidth]{fig1.pdf}}
\centerline{\includegraphics[width=8.7cm]{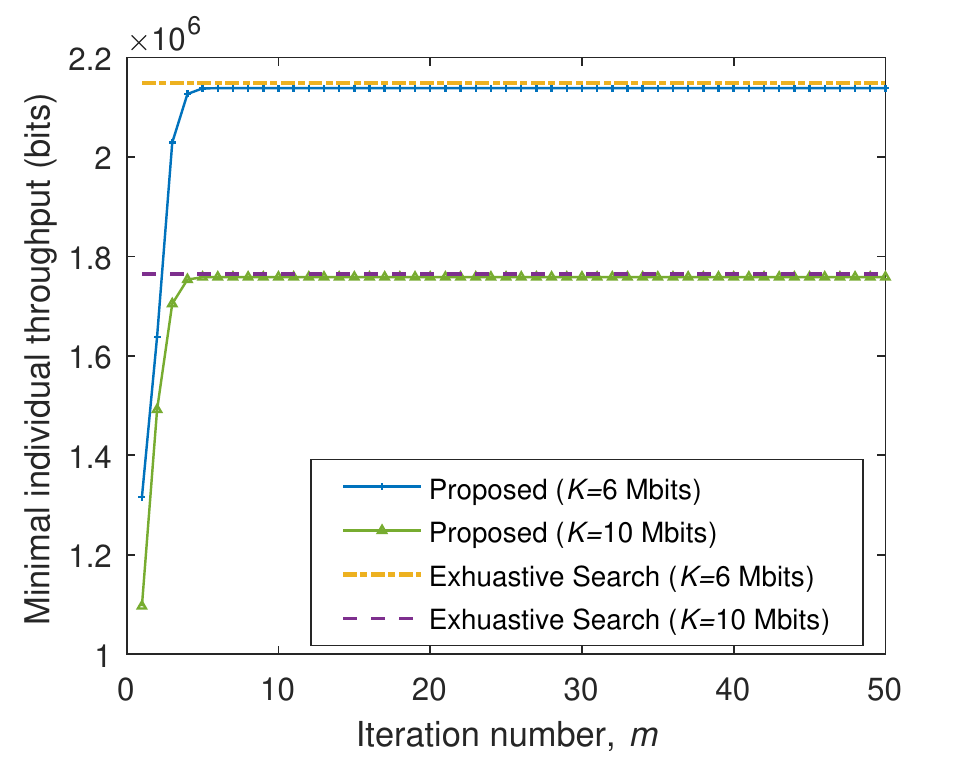}}
\caption{Convergence results and performance gap between the proposed method
and optimal values ($E_{n,\rm{max}}=0.2\,\rm J$ for all $n$, $N=2$).}
\label{fig2}
\end{figure} 
%\begin{figure}[!t]
%%\centerline{\includegraphics[width=\columnwidth]{fig1.pdf}}
%\centerline{\includegraphics[width=8.7cm]{newFig3.pdf}}
%\caption{Convergence of the inner loop of Algorithm 1 for different MDs ($E_{n,\rm{max}}=0.2\,\rm J$ for all $n$, $K=6\,\rm{Mbits}$, $N=4$). Note that this figure is provided on the basis of one random sample.}
%\label{fig3}
%\end{figure}

In Fig. 2, we can see the performance gap between the proposed scheme and the optimal value as well as the convergence performance of the proposed scheme. In order to attain optimal values, we perform the exhaustive search. As shown in the figure, the performance of proposed scheme converges to a stationary point with a small optimality gap in a few iterations, which verifies that the proposed scheme can achieve a near-optimal performance.
%\footnote{The effect of different settings of the required bits $K$ of the common data will be discussed in the analysis about Fig. 4.} 
%Then, similar to \cite{b8}, we take the multiplier $\pmb{\nu}=\{\nu_n\}$ as an example to display the convergence of the inner loop of the Algorithm 1 in Fig. 3. As we can see, the value of $\pmb{\nu}$ converges within the limited steps, which may reflect the convergent property of the subgradient method. Therefore, Algorithm 1 has low computational complexity. 

%\footnote{In Fig. 3, note that the performance of the system follows the performance of the $4$-th MD, whose throughput is the lowest among all MDs in the  example; thus its multipliers remain unchanged.}

%In addition, we also observe that, from Fig.2, for a given requirement of common data amount, such as $K=6M$, the minimal throughput performance decreases as the device number $N$ increases. This is because it is harder for the system to support more MDs to execute the individual data offloading. Although the time consumption in the first stage maybe reduced due to the participation of more MDs, each MD needs to combat the more severe interference among MDs, which is the main factor leading to the decreased minimal throughput of the system. On the contrary, the minimal throughput of the system decreases as the number of the requirement of shared data amount increases for a given device number $N$, since that it is more difficult for meeting the increased requirement of shared data amount, which will directly result in fewer resources that can be used for the second stage. 
\begin{figure}[!t]
%\centerline{\includegraphics[width=\columnwidth]{fig1.pdf}}
\centerline{\includegraphics[width=8.7cm]{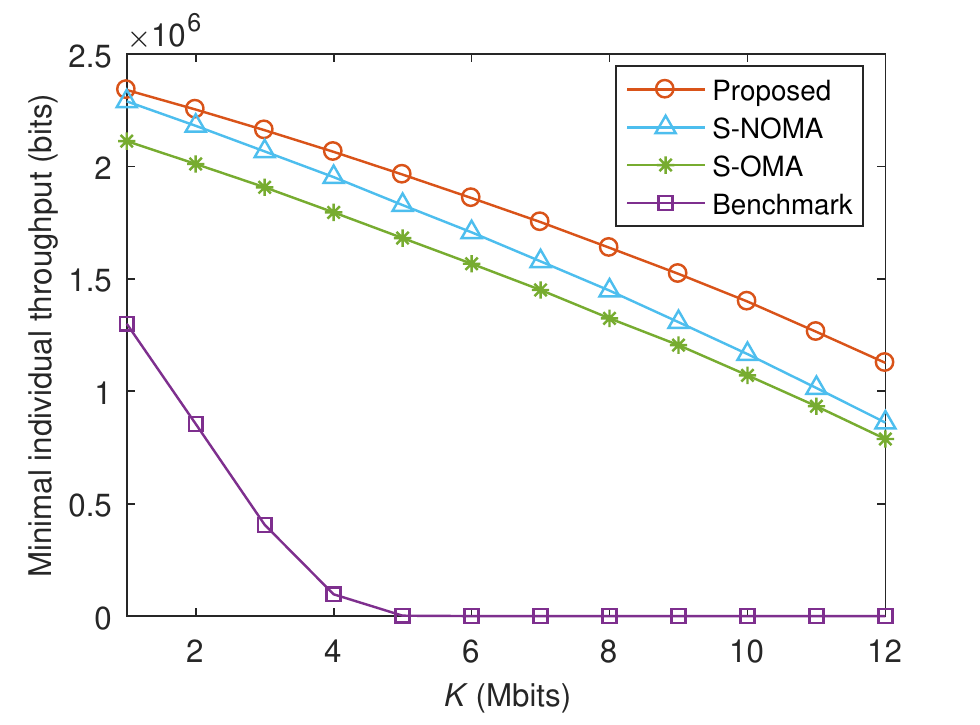}}
\caption{The minimal throughput of the individual data offloading relative to the required bits $K$ of the common data ($E_{n,\rm{max}}=0.2\,\rm J$ for all $n$, $N=4$).}
\label{fig4}
\end{figure}

Fig. 3 shows the performance of the minimal individual offloading throughput of different schemes (the proposed, S-NOMA, S-OMA, and Benchmark) relative to the required bits $K$ of the common data. For each curve, as $K$ increases, the objective performance degrades, because in order to satisfy the requirement of the increased common data amount, both the time and energy consumption in the first stage will be increased, which will directly result in fewer resources available in the second stage. Nevertheless, the advantages of the proposed scheme over the other three schemes (S-NOMA, S-OMA, and Benchmark) are significant, especially when $K$ is large. Although we see that the performance of S-NOMA gets close to our proposed scheme when $K$ is small, as $K$ increases, the performance gap beween the proposed scheme and the two schemes (S-NOMA and S-OMA) is becoming apparent. This is because both S-NOMA and S-OMA employ a single device to offload the entirety of the common data, and this consumes most of the time of system and the energy of the selected device, which results in the relatively poor performance due to the less remaining time and energy. The performance of the Benchmark scheme is always the worst and with the increase of $N$, it declines the fastest compared with the other three schemes. This is because each MD does not distinguish between common data and individual data, and the redundant offloading of the common data by all MDs leads to fewer resources available for offloading the individual data. Furthermore, when $K$ is greater than 5 Mbits, the common data offloading cannot even be maintained, so the performance is always 0. By contrast, our proposed scheme benefits from the advantage of the multi-device cooperation. Since all MDs participate in offloading the common data cooperatively and flexibly  according to its conditions, the proposed scheme can not only avoid redundant offloading for the common data, but also reduce the burden of single device offloading, and accordingly save more resources for the second stage to improve the objective performance.

\begin{figure}[!t]
%\centerline{\includegraphics[width=\columnwidth]{fig1.pdf}}
\centerline{\includegraphics[width=8.7cm]{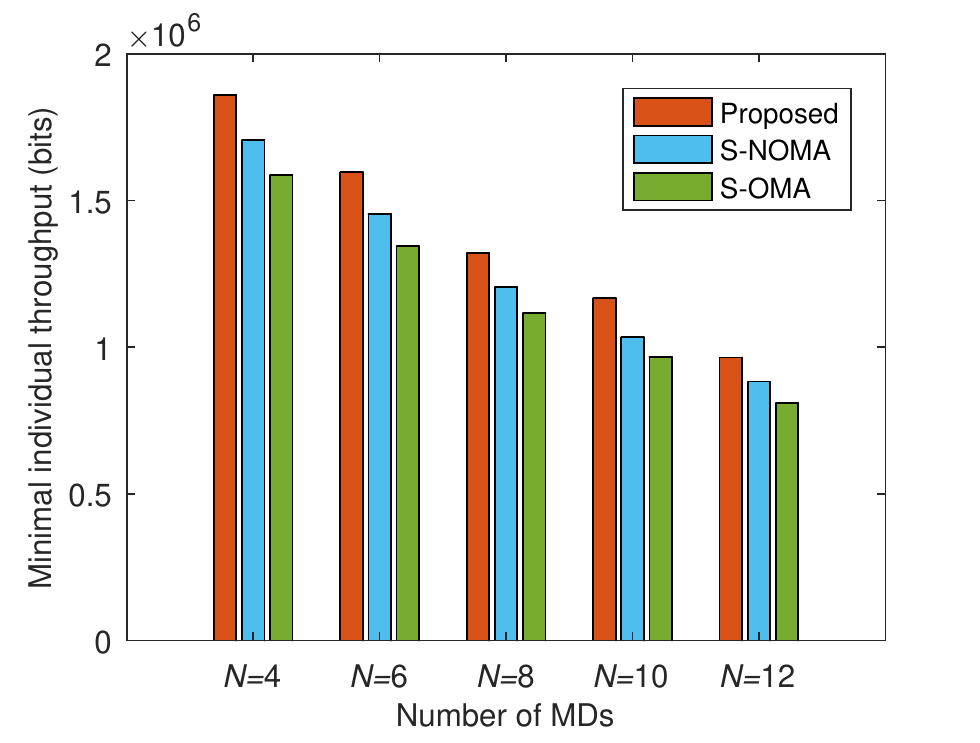}}
\caption{The minimal throughput of the individual data offloading stage relative to the number of mobile devices $N$ ($E_{n,\rm{max}}=0.2\,\rm{J}$ for all $n$, $K=6\,\rm{Mbits}$).}
\label{fig5}
\end{figure}
  
\begin{figure}[!t]
%\centerline{\includegraphics[width=\columnwidth]{fig1.pdf}}
\centerline{\includegraphics[width=8.7cm]{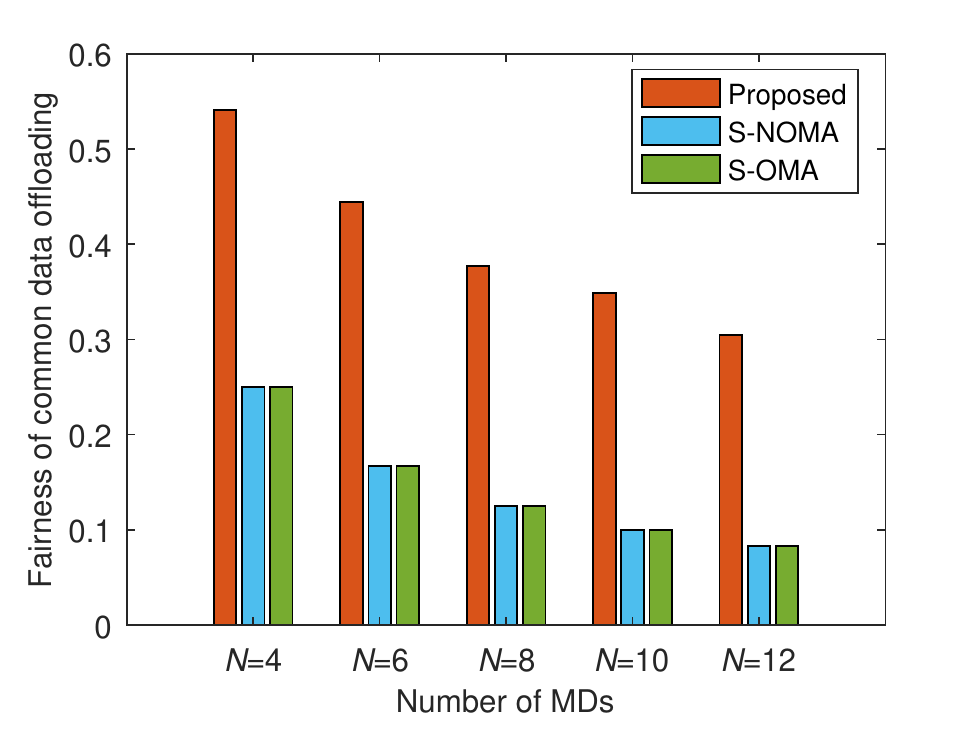}}
\caption{The fairness of the common data offloading stage relative to the number of mobile devices $N$ ($E_{n,\rm{max}}=0.2\,\rm J$ for all $n$, $K=6\, \rm{Mbits}$).}
\label{fig6}
\end{figure}

\begin{figure}[!t]
%\centerline{\includegraphics[width=\columnwidth]{fig1.pdf}}
\centerline{\includegraphics[width=8.7cm]{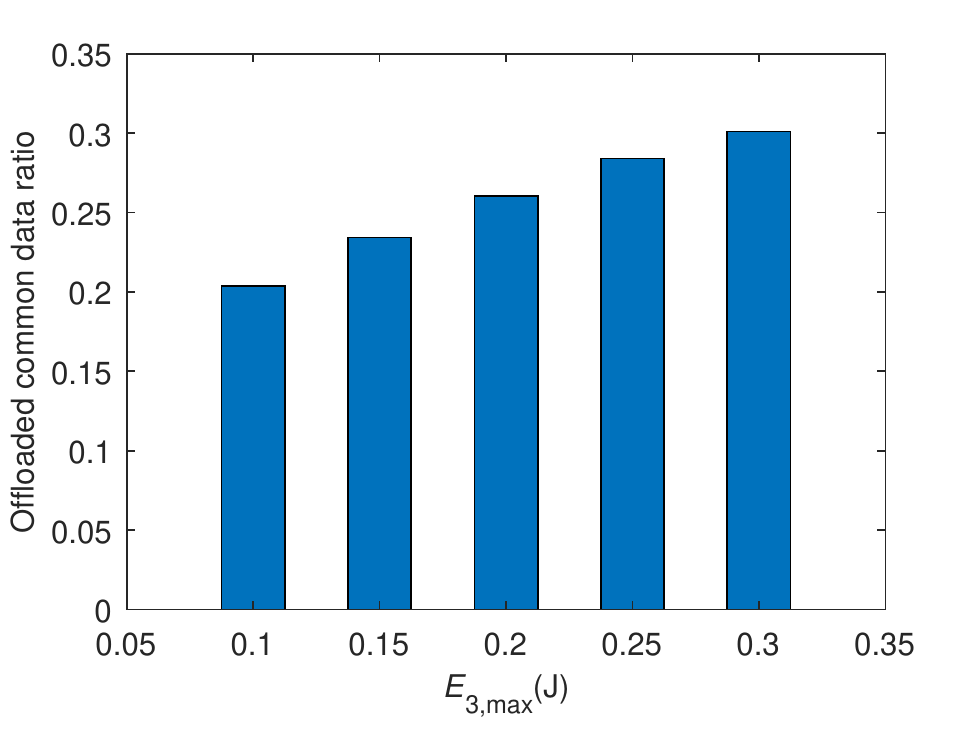}}
\caption{The offloaded data ratio of the $3^{rd}$ MD in common data offloading stage relative to $E_{3,\rm{max}}$ ($E_{n,\rm{max}}=0.2\,\rm{J}$ for $n\in\{1,2,4\}$, $K=6\,\rm{Mbits}$).}
\label{fig7}
\end{figure} 

\begin{figure}[!t]
%\centerline{\includegraphics[width=\columnwidth]{fig1.pdf}}
\centerline{\includegraphics[width=10cm]{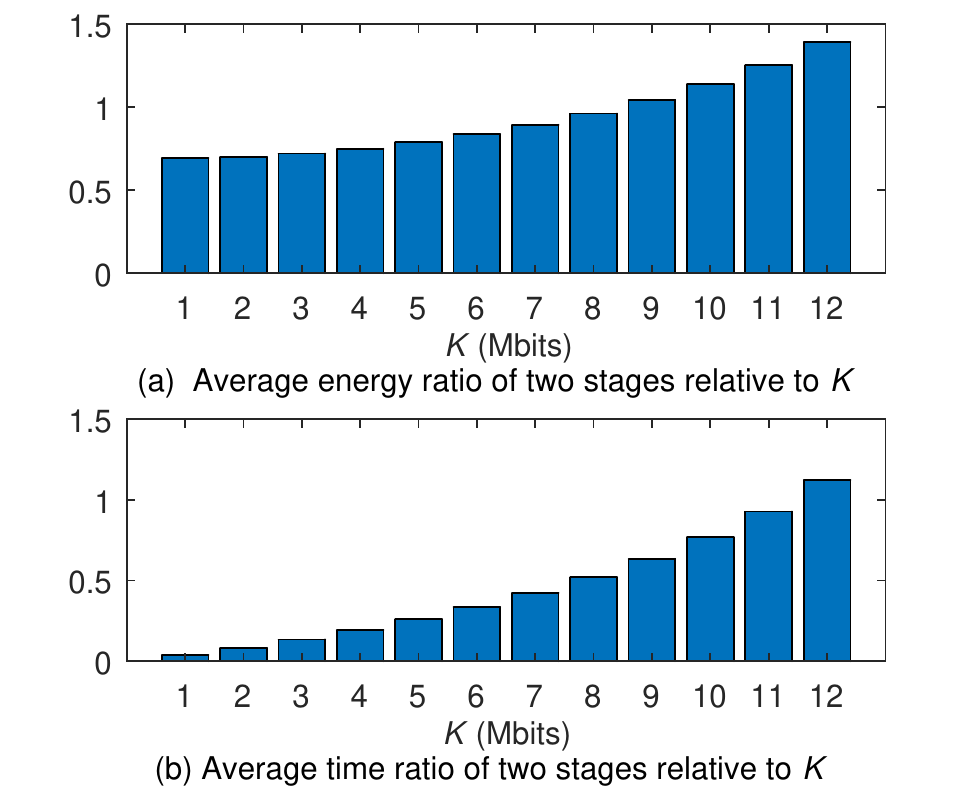}}
\caption{Average energy and time ratio of two stages relative to the required bits $K$ of the common data ($E_{n,\rm {max}}=0.2\,\rm J$ for all $n$, $N=4$).}
\label{fig8}
\end{figure} 
Fig. 4 depicts the performance of the minimal individual offloading throughput of the three schemes (Proposed, S-NOMA, S-OMA)\footnote{The Benchmark scheme is not presented since its performance is always zero under the parameter settings in Fig. 4.} relative to the number of MDs. Overall, for all schemes, the objective performance decreases as $N$ increases. This is because it is harder for the system to support an increasing number of MDs. Nevertheless, benefiting from multi-device collaboration, the proposed scheme is always better than the other two schemes, since in the proposed scheme the scheduling is more flexible in terms of system time allocation and power control of all MDs for the two stages.

Fig. 5 compares the fairness of the three schemes in the first stage with different numbers of MDs, which reflects the fairness experience among MDs in offloading the common data. The fairness is measured by Jain's equation, defined by $\mathcal{J}=\frac{{{{\left( {\sum\limits_{n = 1}^N {{x_n}} } \right)}^2}}}{{N \cdot \sum\limits_{i = 1}^N {{x_n}^2} }}$, where $0\le\mathcal{J}\le 1$. It is apparent that, from Fig. 5, the fairness of the proposed scheme always outperforms the other two schemes. This is because compared with the schemes which employ a single device to offload the entirety of the common data while all MDs share the results, in the proposed scheme, all MDs can participate in offloading the common data, which can improve the fairness among MDs.

In Fig. 6, we take the $3$-rd MD with the setting of available energy $E_{3,\rm{max}}$ ranges from 0.1 J to 0.3 J as an example to illustrate the effect of the different settings of available energy for a single MD on its contribution to the common data offloading. As we can see in Fig. 6, as $E_{3,\rm{max}}$ increases, the volume it offloads in the first stage increases. This indicates that the proposed scheme can feasibly schedule the energy partition according to the energy conditions of MDs. When the MD has less energy, it will contribute less to offloading the common data to ensure the performance of the individual data offloading. When the MD has more energy, it will make a greater contribution to offload the common data. Such a flexible scheduling takes the channel quality and available energy of the MDs into account, and can make reasonable decisions on time allocation and energy partition for the two stages. 

Fig. 7 (a) and (b) demonstrate the average energy ratio $\frac{E_n^C}{E_n^I}$ and time ratio $\frac{\tau^C}{\tau^I}$ of two stages relative to the required bits $K$ of the common data. It is clear that, as $K$ increases, the average energy ratio and time ratio of the two stages also increase. This is because the higher the common data throughput demand, the more time and energy will be consumed in the first stage.

\section{Conclusion}
In this paper, we studied an application-driven computation offloading scheme in a NOMA-enabled MEC network. The scheme was designed to work in two stages for offloading common data and individual data, where multi-device collaboration was utilized to offload the common data. Specifically, the time allocation and power control were jointly optimized to maximize the minimal individual offloading throughput. Utilizing the SCA method, the problem can be solved iteratively. Numerical results verified the advantages of the proposed scheme in achieving better individual offloading performance and improving the fairness experience for the common data offloading. 

Although the offloading scheme in the NOMA-enabled MEC network was deeply investigated, the offloading problem of application-driven multi-server multi-carrier scenarios with jointly optimized user scheduling, sub-carrier allocation, and power control can be pursued in future work. Since traditional optimization may not be able to efficiently cope with the dynamic network environment, offloading schemes should be carried out in systems with the learning capabilities. This will be studied in our future work.

% if have a single appendix:
%\appendix[Proof of the Zonklar Equations]
% or
%\appendix  % for no appendix heading
% do not use \section anymore after \appendix, only \section*
% is possibly needed

% use appendices with more than one appendix
% then use \section to start each appendix
% you must declare a \section before using any
% \subsection or using \label (\appendices by itself
% starts a section numbered zero.)
%

\appendices
\section{Proof of Proposition 1}
1) Assume that when all constraints in \eqref{op5e} are met with equalities, i.e., ${S_{2,n}^*} = \sum\limits_{j = n}^N {E_j^I{\gamma _j}},\forall n\in \mathcal{N}$, the corresponding objective value of problem \eqref{op5} is $\phi_{(2)}^*$. If $\exists n\in \mathcal{N}, S_{2,n}< \sum\limits_{j = n}^N {E_j^I{\gamma _j}}={S_{2,n}^*}$, and the corresponding objective value is $\phi_{(2)}$, it can be deduced that $\phi_{(2)}<\phi_{(2)}^*$ since the left hand of the constraint \eqref{op5c} is monotonously increasing with $S_{2,n}$. Therefore, without loss of optimality, all constraints in \eqref{op5e} will be met with equalities.

2) Similar to the proof in 1), we can obtain the same conclusion about \eqref{op5f}.

3) As for the constraint \eqref{op5d}, the anaysis processing is somewhat more complicated. Also assume when constraint in \eqref{op5d} is met with equality, i.e., ${S_{1}^*}= \sum\limits_{n = 1}^N {E_n^C{\gamma _n}}$, the corresponding objective value of problem \eqref{op5} is $\phi_{(1)}^*$. If when ${S_1}<\sum\limits_{n = 1}^N {E_n^C{\gamma _n}}={S_{1}^*}$, the corresponding objective value of problem \eqref{op5} is $\phi_{(1)}$, then, whether the claim about constraint \eqref{op5d} holds in Proposition 1 will depend on the relationship between $\phi_{(1)}^*$ and $\phi_{(1)}$. 

In this case, we will provide the details of the proof in the following. Based on the results of the optimization when ${S_1}<\sum\limits_{n = 1}^N {E_n^C{\gamma _n}}$ and objective value is $\phi_{(1)}$, if we increase ${S_1}$ to ${S_{1}^*}$, the gap of constraint \eqref{op5b} will also increase, meaning that the total throughput achieved in the first stage will be greater than the required bits $K$. The left hand of the constraint \eqref{op5b} can be defined by 
\begin{equation}\small
f(\tau^C)={\tau ^{ C}}W\log_2 \left(1 + {\frac{S_1}{{{\tau ^{ C}}}}}  \right).\label{f}
\end{equation} 
The first-order derivative of \eqref{f} is calculated as
\begin{equation}\small
f'\left(\tau^C\right){\rm{ = }}\frac{{ W}}{{\ln 2}}\left( {\ln \left( {1 + \frac{{{S_1}}}{{{\tau ^C}}}} \right) - \frac{{\frac{{{S_1}}}{{{\tau ^C}}}}}{{1 + \frac{{{S_1}}}{{{\tau ^C}}}}}} \right).
\end{equation}
 It is difficult to observe the monotonicity according to $f'\left(\tau^C\right)$. Therefore, we further calculate the second-order derivative of $f(\tau^C)$:
\begin{equation}\small
f''\left(\tau^C\right){\rm{ = }}\frac{\rm{ - }\frac{W({S_1})^2}{\left({\tau^C}\right)^3}}{\ln{2}\cdot\left(1+\frac{S_1}{\tau^C}\right)^2}.
\end{equation}
%$\frac{S_1}{\tau^C}$ is approaching to $0^+$,
It is clear that $f''\left(\tau^C\right)<0$, which indicates $f'\left(\tau^C\right)$ is monotonously decreasing. When $\tau^C \to \infty$ \footnote{Actually, the maximal value of $\tau^C$ is $T_{\max}$ and we just use $\tau^C \to \infty $ to verify the value range of $f'(\tau^C)$.}, $\ln\left(1+\frac{S_1}{\tau^C}\right)\sim \frac{S_1}{\tau^C}$ holds on the basis of the principle of the equivalent infinitesimal. Consequently, $f'\left(\tau^C\to \infty\right)$=$\frac{\left(\frac{S_1}{\tau^C}\right)^2}{1+\frac{S_1}{\tau^C}}\Bigg|_{\tau^C\to \infty}$ is approaching $0^+$. Combining the property of monotonous decreasing of $f'\left(\tau^C\right)$, we obtain $f'\left(\tau^C\right)>0$, wherein $\tau^C \in (0,T_{\max}]$. Thus, we can determine that $f(\tau^C)$ is a monotonously increasing function, meaning that we can decrease $\tau^C$ to make \eqref{op5b} an equality.

Next, we check the monotonicity of the left hand of \eqref{op5c} and \eqref{op5g}. According to the discussion of $f(\tau^C)$, we learn that the left hand of \eqref{op5g}, which can be expressed by $o(\tau^I)={\tau ^{ I}}W{\log _2}\left(1 + { \frac{S_{2,N}}{{{\tau ^{ I}}}}}\right ) $, is also a monotonuously increasing function of $\tau^I$. As for the left hand of \eqref{op5c}, it can be written as 
\begin{equation}\small
\begin{split}
g(\tau^I)=&{\tau ^{ I}}W{\log _2}\left(1 + { \frac{S_{2,n}}{{{\tau ^{ I}}}}}\right ) -{\tau ^{ I}}W{\log _2}\left(1 + { \frac{S_{3,n}}{{{\tau ^{ I}}}}}\right )\\
=&\frac{\tau^IW}{\ln2}\ln\left(\frac{\tau^I+S_{2,n}}{\tau^I+S_{3,n}}\right).
\end{split}
\end{equation}
We calculate the first-order derivative of $g(\tau^I)$ as follows:
\begin{equation}\small\label{gfirst}
g'(\tau^I)=\frac{W}{\ln2}\left(\ln\left(\frac{\tau^I+S_{2,n}}{\tau^I+S_{3,n}}\right)+\frac{\tau^I(S_{3,n}-S_{2,n})}{(\tau^I+S_{2,n})(\tau^I+S_{3,n})}\right).
\end{equation}
Note that from the above proofs 1) and 2) for Proposition 1, ${S_{2,n}} = \sum\limits_{j = n}^N {E_j^I} {\gamma _j}$ and ${S_{3,n}} = \sum\limits_{k = n + 1}^N {E_k^I} {\gamma _k}$ hold when problem \eqref{op5} achieves its optimality. Hence, we can obtain $S_{2,n}\ge S_{3,n}, \forall n\in\{1,\ldots,N-1\}$. Observing $g'(\tau^I)$, we cannot see the monotonicity directly. Therefore, we calculate the second-order derivative, which is given by
\begin{equation}\small 
g''(\tau^I)=\dfrac{W(S_{3,n}-S_{2,n})(S_{2,n}\tau^I+S_{3,n}\tau^I+2S_{2,n}S_{3,n})}{\ln2\cdot{(\tau^I+S_{2,n})^2(\tau^I+S_{3,n})}^2}.
\end{equation} 
It is clear that $g''(\tau^I)\le 0$, which means that $g'(\tau^I)$ is also monotonously decreasing. Namely, we can obtain $g'(\tau^I)$ satisfies that 
\begin{equation}\small\label{gfirstrange}
g'(\tau^I\in(0, T_{max}])\ge g'(\tau^I\to \infty).
\end{equation}
When $\tau^I\to \infty$, \eqref{gfirst} can be expressed by 
\begin{equation}\small
\begin{split}
&\mathop {\lim }\limits_{\tau^I \to \infty } g'(\tau^I)\\
=&\frac{W}{\ln2}\left(\ln\left(\frac{\tau^I+S_{2,n}}{\tau^I+S_{3,n}}\right)+\frac{\tau^I(S_{3,n}-S_{2,n})}{(\tau^I+S_{2,n})(\tau^I+S_{3,n})}\right)\Bigg{|}_{\tau^I\to \infty }\\
=&\frac{W}{\ln2}\left(\ln\left(1+\dfrac{S_{2,n}-S_{3,n}}{\tau^I+S_{3,n}}\right)+\frac{\tau^I(S_{3,n}-S_{2,n})}{(\tau^I+S_{2,n})(\tau^I+S_{3,n})}\right)\Bigg{|}_{\tau^I\to \infty }\\
=&\frac{W}{\ln2}\left(\dfrac{S_{2,n}-S_{3,n}}{\tau^I+S_{3,n}}+\dfrac{\tau^I(S_{3,n}-S_{2,n})}{(\tau^I+S_{2,n})(\tau^I+S_{3,n})}\right)\Bigg{|}_{\tau^I\to \infty }\\
=&\frac{W}{\ln2}\left(\frac{S_{2,n}(S_{2,n}-S_{3,n})}{(\tau^I+S_{2,n})(\tau^I+S_{3,n})}\right)\Bigg{|}_{\tau^I\to \infty }\\
=&0^+.
\end{split}
\end{equation}
wherein $\ln\left(1+\frac{S_{2,n}-S_{3,n}}{\tau^I+S_{3,n}}\right)\sim\frac{S_{2,n}-S_{3,n}}{\tau^I+S_{3,n}}$ holds when $\tau^I\to \infty$ based on the principle of the equivalent infinitesimal.

%, $\frac{S_{2,n}-S_{3,n}}{\tau^I+S_{3,n}}$ is approaching $0^+$,

From \eqref{gfirstrange}, we know that $g'(\tau^I)\ge 0$. Accordingly, $g(\tau^I)$ is also monotonously increasing. The monotonicity of $o(\tau^I)$ and $g(\tau^I)$ indicates that when we decrease $\tau^C$ to make constraint \eqref{op5b} an equality, we can increase $\tau^I$ within the time constraint \eqref{op1b} to increase $o(\tau^I)$ in constraint \eqref{op5g} and $g(\tau^I)$ in constraint \eqref{op5c}, accordingly improving the objective value. That is to say, $\phi_{(1)}^*>\phi_{(1)}$ holds. Thus, we can conclude that without loss of optimality to problem \eqref{op5}, constraint \eqref{op5d} will be met with equality. 
% you can choose not to have a title for an appendix
% if you want by leaving the argument blank
\section{Proof of Proposition 2}
The left hand of \eqref{op5b} is given by 
\begin{equation}\small\label{fjoint}
f(\tau^C, S_1)={\tau ^{ C}}W\log_2 \left(1 + {\frac{S_1}{{{\tau ^{ C}}}}}  \right),
\end{equation} 
and the left hand of \eqref{op5c} is given by
\begin{equation}\small\label{gjoint}
\begin{split}
g(\tau^I, S_{2,n}, S_{3,n})=&{\tau ^{ I}}W{\log _2}\left(1 + { \frac{S_{2,n}}{{{\tau ^{ I}}}}}\right ) -{\tau ^{ I}}W{\log _2}\left(1 + { \frac{S_{3,n}}{{{\tau ^{ I}}}}}\right )\\
=&g_1(\tau^I, S_{2,n})-g_2(\tau^I, S_{3,n}).
\end{split}
\end{equation}
Observing \eqref{fjoint} and \eqref{gjoint}, we see that the functions of $f(\tau^C, S_1)$, $g_1(\tau^I, S_{2,n})$ and $g_2(\tau^I, S_{3,n})$ have a similar form, which can be expressed as follows:
\begin{equation}\small\label{joint}
q\left(a, b\right)={a}\log_2\left(1 + {\frac{b}{{a}}}  \right).
\end{equation} 
The second-order Hessian Matrix of \eqref{joint} in variables $a$ and $b$ is calculated as follows:
\begin{equation}\small
\mathcal{H}{\rm{ = }}\frac{1}{{\ln 2 \cdot a{{(1 + \frac{b}{a})}^2}}}\Bigg( {\begin{array}{*{20}{c}}
{ - \frac{{{b^2}}}{{{a^2}}}}&{\frac{b}{a}}\\
{\frac{b}{a}}&{ - 1}
\end{array}} \Bigg).
\end{equation}
Note that $\mathcal{H}$ is a symmetric matrix. Furthermore, if $\forall{t}\in{\mathbb{R}^n}$, matrix $\mathcal{H}$ satisfies the formula $t^*Ht\le 0$, we can conclude that $\mathcal{H}$ is a negative semi-definite matrix. As a consequence, $q\left(a, b\right)$ is jointly concave in variables $a$ and $b$.

According to the concavity of $q\left(a, b\right)$, we know that $f(\tau^C, S_1)$, $g_1(\tau^I, S_{2,n})$ and $g_2(\tau^I, S_{3,n})$ are concave functions in their correlated variables, thereby illustrating the concavity of the left hand of constraint \eqref{op5b} as well as the D.C. form of constraint \eqref{op5c}. 
\section{Proof of Proposition 3}
In order to prove that the desired objective values in a sequence of approximated problems are non-decreasing, we need to verify that the solution of problem \eqref{op6} in the $(m-1)$-th iteration is always feasible in the $m$-th iteration.

In the $m$-th iteration, the optimal solution of problem \eqref{op5} is assumed to be $\Lambda=\{\phi^*,{\tau}^{{ C}{*}},{\tau}^{ I*},\pmb{E^{ C}}^*,\pmb{E^{ I}}^*,{S_1}^*,\pmb{S_2}^*,\pmb{S_3}^*\}$.  Here, the corresponding expression satisfied by the optimal solution in the approximated constraint \eqref{op6b} is given by  
\begin{equation}\small\label{c48}
\begin{split}
&{\tau ^{ I*}}W{\log _2}\left(1 + { \frac{S_{2,n}^*}{{{\tau ^{ I*}}}}}\right )-B_n\Big(\tau^I[m-1], S_{3,n}[m-1]\Big)\\
-&D_n\Big(\tau^I[m-1], S_{3,n}[m-1]\Big)\Big(S_{3,n}^*-S_{3,n}[m-1]\Big)\\
-&Q_n\Big(\tau^I[m-1], S_{3,n}[m-1]\Big)\Big(\tau ^{I*}-\tau ^I[m-1]\Big)\ge \phi^* ,\ \forall n.
\end{split}
\end{equation}
Then, in the $m$-th iteration, we use the optimal solution obtained in the $(m-1)$-th iteration to update the local point, i.e., $\tau^I[m]=\tau^{I*}$, and $S_{3,n}[m]=S_{3,n}^*$. Substituting the optimal solution $\Lambda$ in constraint \eqref{op5b} with the updated parameter, we get the following result:
\begin{subequations}\small
\begin{align}
&{\tau ^{ I*}}W{\log _2}\left(1 + { \frac{S_{2,n}^*}{{{\tau ^{ I*}}}}}\right )-B_n\Big(\tau^{I*}, S_{3,n}^*\Big)\nonumber\\
&-D_n\Big(\tau^{I*}, S_{3,n}^*\Big)\Big(S_{3,n}^*-S_{3,n}^*\Big)-Q_n\Big(\tau^{I*}, S_{3,n}^*\Big)\Big(\tau ^{I*}-\tau ^{I*}\Big)\label{49a}\\
&= {\tau ^{ I*}}W{\log _2}\left(1 + { \frac{S_{2,n}^*}{{{\tau ^{ I*}}}}}\right )-{\tau ^{ I*}}W{\log _2}\left(1 + { \frac{S_{3,n}^*}{{{\tau ^{ I*}}}}}\right )\label{49b}\\
&\ge{\tau ^{ I*}}W{\log _2}\left(1 + { \frac{S_{2,n}^*}{{{\tau ^{ I*}}}}}\right )-B_n\Big(\tau^I[m-1], S_{3,n}[m-1]\Big)\nonumber\\
&-D_n\Big(\tau^I[m-1], S_{3,n}[m-1]\Big)\Big(S_{3,n}^*-S_{3,n}[m-1]\Big)\nonumber\\
&-Q_n\Big(\tau^I[m-1], S_{3,n}[m-1]\Big)\Big(\tau ^{I*}-\tau ^I[m-1]\Big)\label{49c}\\
&\ge\  \phi^* ,\ \forall n.\label{49d}
\end{align}
\end{subequations}
Since the last two terms in \eqref{49a} are 0, we obtain the expression \eqref{49b}. Furthermore, due to its concavity, the second term of \eqref{49b} is upper-bounded by its first-order Taylor approximation in the $(m-1)$-th iteration, i.e.,
\begin{equation}\small\label{gradient}
\begin{split}
B_n&\Big(\tau^{I*}, S_{3,n}^*\Big)={\tau ^{ I*}}W{\log _2}\left(1 + { \frac{S_{3,n}^*}{{{\tau ^{ I*}}}}}\right )\\
\le &B_n\Big(\tau^I[m-1], S_{3,n}[m-1]\Big)\\
&+D_n\Big(\tau^I[m-1], S_{3,n}[m-1]\Big)\Big(S_{3,n}^*-S_{3,n}[m-1]\Big)\\
&+Q_n\Big(\tau^I[m-1], S_{3,n}[m-1]\Big)\Big(\tau ^{I*}-\tau ^I[m-1]\Big).
\end{split}
\end{equation} 
From \eqref{gradient}, it can be deduced that \eqref{49c} holds. Comparing with \eqref{c48}, we get the resulting expression in \eqref{49d}. Hence, the solution $\Lambda=\{\phi^*,{\tau}^{{ C}{*}},{\tau}^{ I*},\pmb{E^{ C}}^*,\pmb{E^{ I}}^*,{S_1}^*,\pmb{S_2}^*,\pmb{S_3}^*\}$ in the $(m-1)$-th iteration is always a feasible point for problem \eqref{op5} in the $m$-th iteration.

Since problem \eqref{op6} is a concave problem, the objective value in the $m$-th iteration is either unchanged or improved compared with the value in the $(m-1)$-th iteration. With the application of the SCA method, the improved solution is always applied as the local point in the next iteration to further improve the resulting solution, yielding the convergence to a stationary point, i.e., the local maximum is obtained. 
% use section* for acknowledgment
%\section*{Acknowledgment}

%The authors would like to thank...

\ifCLASSOPTIONcaptionsoff
  \newpage
\fi

% trigger a \newpage just before the given reference
% number - used to balance the columns on the last page
% adjust value as needed - may need to be readjusted if
% the document is modified later
%\IEEEtriggeratref{8}
% The "triggered" command can be changed if desired:
%\IEEEtriggercmd{\enlargethispage{-5in}}

% references section

% can use a bibliography generated by BibTeX as a .bbl file
% BibTeX documentation can be easily obtained at:
% http://mirror.ctan.org/biblio/bibtex/contrib/doc/
% The IEEEtran BibTeX style support page is at:
% http://www.michaelshell.org/tex/ieeetran/bibtex/
%\bibliographystyle{IEEEtran}
% argument is your BibTeX string definitions and bibliography database(s)
%\bibliography{IEEEabrv,../bib/paper}
%
% <OR> manually copy in the resultant .bbl file
% set second argument of \begin to the number of references
% (used to reserve space for the reference number labels box)
%\begin{thebibliography}{1}
%
%\bibitem{IEEEhowto:kopka}
%H.~Kopka and P.~W. Daly, \emph{A Guide to \LaTeX}, 3rd~ed.\hskip 1em plus
%  0.5em minus 0.4em\relax Harlow, England: Addison-Wesley, 1999.
%
%\end{thebibliography}
\bibliographystyle{ieeetr}
\bibliography{IEEEabrv,iotreference}
% biography section
% 
% If you have an EPS/PDF photo (graphicx package needed) extra braces are
% needed around the contents of the optional argument to biography to prevent
% the LaTeX parser from getting confused when it sees the complicated
% \includegraphics command within an optional argument. (You could create
% your own custom macro containing the \includegraphics command to make things
% simpler here.)
%\begin{IEEEbiography}[{\includegraphics[width=1in,height=1.25in,clip,keepaspectratio]{mshell}}]{Michael Shell}
% or if you just want to reserve a space for a photo:

%\begin{IEEEbiography}{Michael Shell}
%Biography text here.
%\end{IEEEbiography}

% if you will not have a photo at all:

%\begin{IEEEbiographynophoto}{Jane Doe}
%Biography text here.
%\end{IEEEbiographynophoto}

%\vfill

% Can be used to pull up biographies so that the bottom of the last one
% is flush with the other column.
%\enlargethispage{-5in}

% that's all folks
\end{document}